# High contrast linking 6 lasers to a 1 GHz Yb:fiber laser frequency comb

Yuxuan Ma[1], Fei Meng[1,2], Yan Wang[1], Aimin Wang[1], and Zhigang Zhang[1*]

[1]*State Key Laboratory of Advanced Optical Communication System and Networks,*
*School of Electronics Engineering and Computer Science, Peking University, Beijing 100871, China*
[2]*National Institute of Metrology, Beijing 100029, China*
*\*Corresponding author: zhgzhang@pku.edu.cn*

We demonstrate a 0.95 GHz repetition rate fully stabilized Yb:fiber frequency comb without optical amplification. Benefitted from the high mode power and high coherence, this comb achieved 35 dB to 42 dB signal to noise ratio on the direct heterodyne beat signals with at least 6 continuous wave lasers (at 580nm, 679nm, 698nm, 707nm, 813nm and 922 nm) while keeping >40 dB carrier envelop frequency signal. It can be used for the direct measurement of optical frequencies in visible and near-infrared wavelengths, and has a great potential on simultaneous comparison of multiple optical frequencies.

In the past decade, optical frequency combs (OFCs) have become a powerful tool bridging radio frequency (RF) and optical frequency. As the OFCs can tightly trace the atomic reference and provide abundant individual optical frequencies, they have been applied in various high-precision experiments including absolute optical frequency measurement [1], absolute distance measurement [2], time and frequency transfer [3], ultra-low noise microwave generation [4], calibration of astronomical spectrograph [5], and comparison between optical clocks [6].

With the development of optical clocks, a growing demand for OFCs is to simultaneously link more optical frequencies, especially those wavelengths in the visible and near infrared range, such as 578 nm (Yb optical lattice clock), 657 nm (Ca atomic clock), 698 nm (Sr optical lattice clock), 1070 nm/4 (Al+ ion clock) and 1126 nm/4 (Hg+ ion clock). A traditional choice for the optical frequency metrology is the Ti:sapphire lasers that feature the low intrinsic noise and broadband mode locking spectrum [1,7]. However, it is difficult for Ti:sapphire laser based combs to achieve long-term operation while to maintain the signal to noise ratio (SNR) high in both carrier envelop offset frequency ($f_{ceo}$) and multiple heterodyne beat signals ($f_{beat}$) with different lasers.

Fiber laser based combs are becoming competitors to Ti:sapphire combs in their self-starting, long-term stable operation, low cost and small foot-print. Recently, single-branch Er:fiber laser based OFCs have been proved a fractional frequency instability to $10^{-18}$ at 1 s for comparison of optical atomic clocks, in addition to their robustness and easy operation [8,9]. However, Er:fiber lasers deliver low average output power (<100 mW) and low pulse energy, so that extra optical amplification stages are required to achieve enough peak pulse power for octave-spanning spectrum generation and frequency conversion. The amplification process may deteriorate the coherence to some extent [10]. Another drawback of Er:fiber based OFCs is that their super-continuum (SC) spectrum (generally from 1μm to 2μm) does not directly cover the visible range, where exists many wavelengths of optical clock lasers. Although frequency doubling could help them to reach the visible range, the resulted SNR of beat signal will become distinctly lower [9].

Previously we have developed a 1 GHz mode locked Yb:fiber laser [11], in which the major part of the pulse energy was used to generate octave-spanning spectrum for obtaining the $f_{ceo}$. The rest of pulse energy was to go for amplification and applications.

In this paper, we propose to use the full output power of a GHz repetition rate Yb:fiber OFC for generating a more functional SC spectrum, which is used for both the high contrast self-referenced $f_{ceo}$ signal and heterodyne beat signals with multiple wavelengths in a single port, without amplifiers. We demonstrate high SNR $f_{beat}$ signals between the comb and 6 narrow linewidth continuous wave (CW) lasers, which are all essential wavelengths in optical clock systems.

The system setup is shown in Fig. 1(a). The seed laser is the same as that in ref. [11]. It delivers a train of near-Fourier-transform-limited pulses (~70 fs) with 600 mW average power at 954 MHz repetition rate. All of the output power is coupled into a piece of tapered photonic crystal fiber (PCF) with ~60% coupling efficiency to generate SC spectrum more than one octave. The SC spectrum will be used to simultaneously generate $f_{ceo}$ and several $f_{beat}$ signals. To keep the high coherence over broad bandwidth, we chose a single mode PCF (SC-4.6-1000-46BB0, YOFC Co., Ltd.) with high nonlinearity and zero dispersion around

1040 nm to allow efficient nonlinear interactions through the Yb:fiber laser output. Following the simulation in Ref. [12], we choose to reduce the pitch of the PCF from 3.1 µm to 1.2 µm by tapering the diameter of middle part from 119 µm down to 46 µm, thus a broadband over one octave can be obtained. As the coherence may be deteriorated by unnecessary long nonlinear fiber, the length of the tapered waist was set to 5 cm, which is long enough for the SC generation without breaking the coherence at the spectrum edges. The near-Fourier-transform-limited 70 fs pulse from laser oscillator is also helpful to maintain the coherence during the SC generation process because of the short pulse duration and low amplified spontaneous emission (ASE) noise [13]. Counting the 2 × 4 cm long untapered parts and 2 × 3 cm long transition parts in, the total length of the PCF is 19 cm. The PCF was spliced to a commercial collimator with very short single mode fiber pigtail to guarantee high and stable coupling efficiency. The structure of this collimator-PCF is shown in Fig. 1(b).

Figure 2(a) shows the spectra of the input pulse and the SC. A photograph of the grating dispersed laser beam is shown in Fig. 2(b) since the optical spectrum analyzer (YOKOGAWA-AQ6375) could not display the spectrum below 600 nm.

For $f_{ceo}$ detection, the traditional f-2f self-referencing method was employed. The SC spectrum was separated into two arms, namely visible arm (<900 nm) and infrared arm (>900 nm), by a dichroic mirror (DM1). The infrared arm was built as a delay line to synchronize the recombined pulses from the two arms to improve the SNR of $f_{ceo}$. After the recombination of the two arms, the laser was sent into in a periodically poled lithium niobate (PPLN) crystal to generate the second harmonic of the 1220 nm wavelength part. The laser beam was then spatially dispersed by a blazed grating (1200 lines/mm). The 0th order diffraction light was received by a photo detector (PD) as the $f_{rep}$ signal, while the 1st order diffraction light was launched into an avalanche photo detector (APD) to obtain $f_{ceo}$ signal by beating the original and frequency doubled 610 nm components (see Fig. 3(c)). Currently the $f_{ceo}$ and $f_{rep}$ are phase locked to RF standard with $10^{-12}\tau^{-1/2}$ frequency instability.

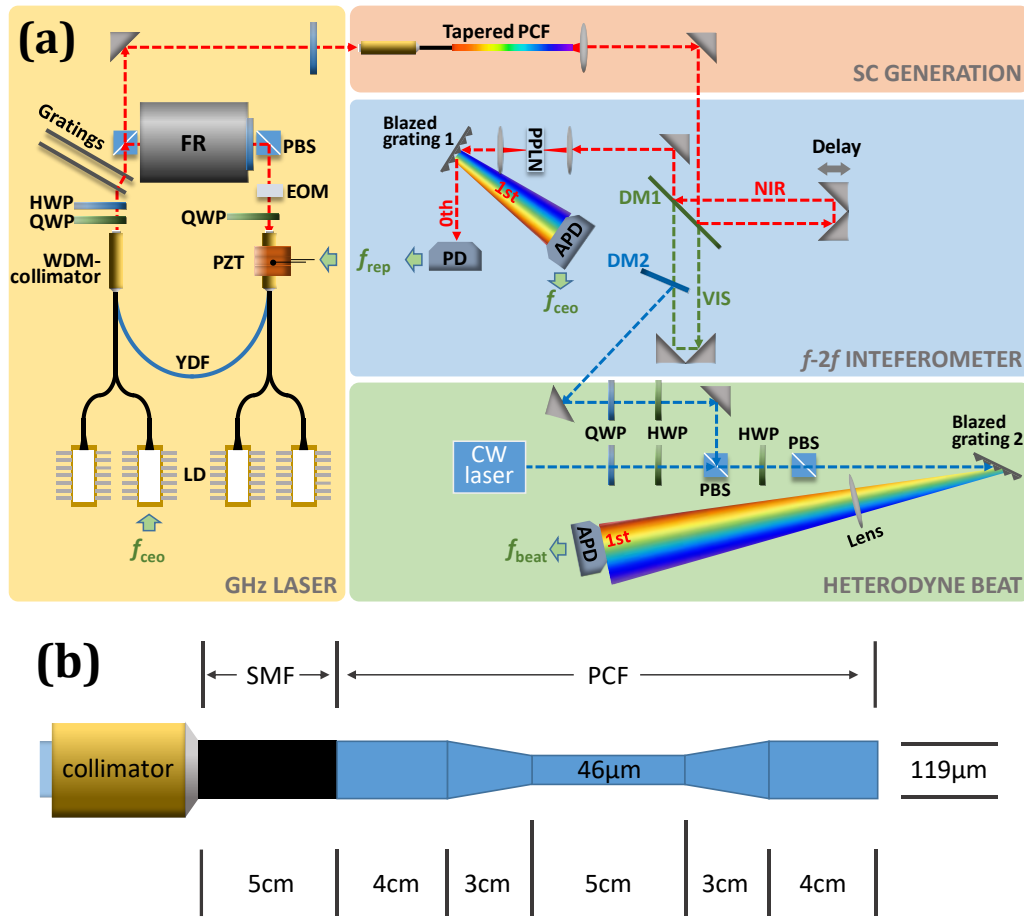

Fig. 1. (a) The configuration of whole setup. HWP: half wave plate; QWP: quarter wave plate; PBS: polarization beam splitter; YDF: ytterbium doped fiber; LD: pump laser diode; FR: Faraday rotator; NIR: near-infrared part; VIS: visible part; (b) The structure of the tapered PCF spliced to fiber collimator.

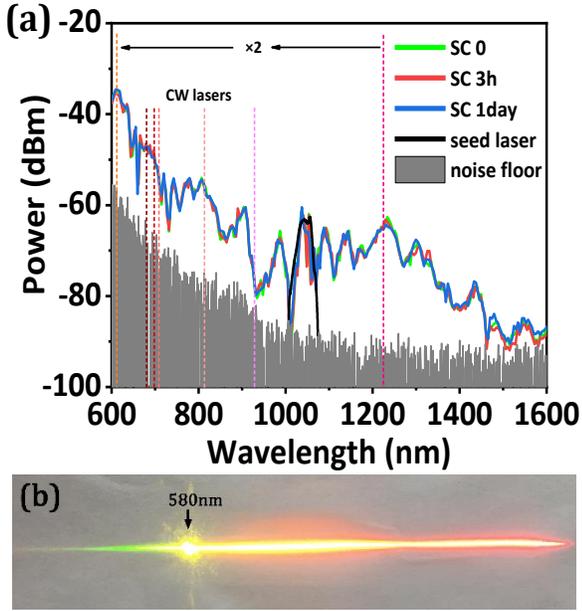

Fig. 2. (a) The mode locking spectrum (black curve) and the generated SC spectra over 0 second (green curve), 3 hours (red curve) and one day (blue curve). The grey part is the noise floor of the optical spectrum analyzer. Those wavelengths used for $f_{ceo}$ and $f_{beat}$ detections are marked by vertical dashed lines. (b) A photo of the spatially dispersed SC laser beam. The yellow spot on the left is the CW laser at 580 nm.

In the visible arm, we inserted another dichroic mirror (DM2) to pick out the desired wavelength for heterodyne beat. By spatial combination and dispersion of this light together with a CW laser operating at corresponding wavelength, the $f_{beat}$ signal was obtained simultaneously with the $f_{ceo}$. Six different CW lasers at 580 nm, 679 nm, 698 nm, 707 nm, 813 nm and 922 nm were respectively used to beat with the comb. Those lasers are all dependent on the availability of the laboratory.

The beat notes of the first five lasers were obtained together with the $f_{ceo}$ signal, whereas the beat note of 922 nm has to block the $f_{ceo}$ due to the lack of suitable DM to pick the wavelength during the experiment was conducted. The RF spectra of the $f_{beat}$ signals are plotted in Fig. 3(a) and 3(b). In particular, the beat with 922 nm CW laser is displayed separately in Fig. 3(b) because the laser had an ineradicable internal side-band modulation at 20 MHz. The linear density of the blazed grating for heterodyne beat was 2400 lines/mm for 580 nm to 707 nm lasers, but 1000 lines/mm for 813 nm and 922 nm lasers. A lens with 200 mm focal length was used to focus the diffracted light, and the optical power of these CW lasers in front of the APD were all around 100 μW.

It is seen that the SNR of the beat notes varied from 35 dB to 42 dB for different wavelengths, meanwhile the SNR of $f_{ceo}$ signal maintained >40 dB, for the resolution bandwidth (RBW) of 100 kHz. Such high contrast signals are favorable for the subsequent frequency counting or phase locking.

We monitored the SC spectrum over 3 hours and one day (the red and the blue curves respectively in Fig. 2(a)), and observed almost no changes. A single beat signal with 580 nm laser was also monitored overnight, and no degradation on SNRs was recognized even after the system restarts without any re-alignment.

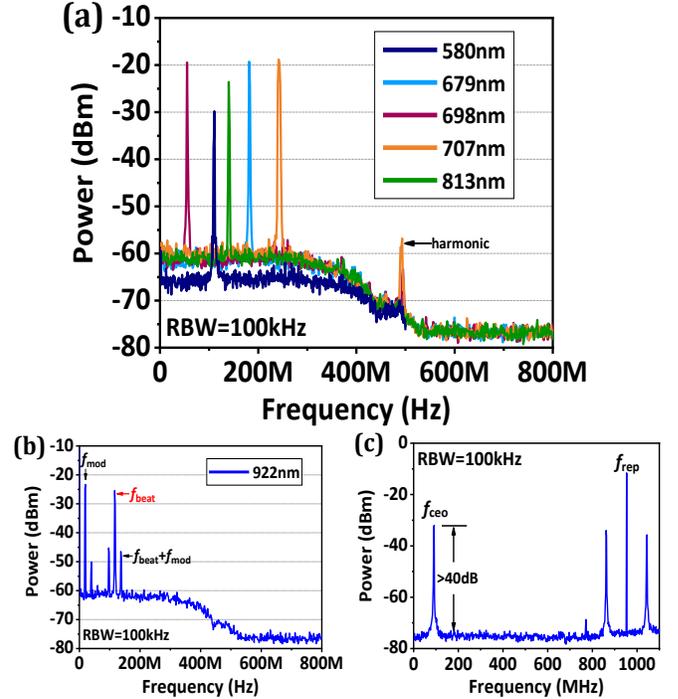

Fig. 3. (a) The free-running heterodyne beat signals between the comb and the CW lasers at 580 nm, 679 nm, 698 nm, 707 nm, and 813 nm; (b) The free-running beat signal between the comb and the 922 nm CW laser; (c) The free-running $f_{ceo}$ signal.

Although we have only tried six different wavelengths, it is very likely that this comb could link all the wavelengths from 580 nm to 1300 nm with similar SNR. All those above involved wavelengths (including the 610 nm and 1220 nm for $f_{ceo}$ detection) have spread over the SC spectrum, and those wavelengths located in the mode locking spectrum, such as 1064 nm, are surely in high SNR [14]. It is remarkable that the spectrum intensity at 922 nm is almost the weakest part within the SC below 1300 nm, whereas the SNR can still be as high as 35 dB. Longer wavelength range up to 1560 nm is also in the coverage of the SC spectrum, in spite of low intensity. Limited by the availability of the filters and the detectors, however, the beat with lasers in the range of 1300 nm to 1560 nm range was not successfully attempted. If several proper filters are used, more wavelengths can be simultaneously measured or locked to the comb.

The high SNR is partially due to the high comb teeth power and the maintaining of coherence, and the stable SC is owing to the collimator spliced to tapered PCF and the good self-starting of the laser. The single port configuration without amplification stages makes the comb inherently away from uncommon fiber path noise between the beat notes and the ASE noise brought by fiber amplifiers. If locked to an optical reference such as an optical clock laser, the comb could inherit the optical frequency stability and linewidth [6], and is able to show ~$10^{-18}$/s fractional frequency instability and <0.5 rad residual phase error with enabling high speed servo control [14].

In summary, we have demonstrated for the first time the high contrast beating signals with a GHz repetition rate Yb:fiber laser frequency comb. Benefitted from the high repetition rate, high output power of the laser, and the home-made tapered PCF, the comb exhibits the high teeth power and high coherence over an octave-spanning in single port, without the need of amplification. The SC spectrum of the comb was directly used to beat with 6 CW lasers at the wavelength of 580 nm, 679 nm, 698 nm, 707 nm, 813 nm, and 922 nm. More than 40 dB $f_{ceo}$ signal and 35 dB ~ 42 dB beat notes at those wavelengths have been obtained, even at the spectral dip around 922 nm. Both the broadband SC and the SNR of the beating signals showed good long-term stability and reproducibility.

The full stabilization to RF or optical standards make the comb a very useful tool for simultaneous optical frequency measurement or stabilization of multiple CW lasers. This simple and compact comb will also find more advanced applications (like optical comparison) in the future.

This research was supported by the National Natural Science Foundation of China (NSFC) (61575004, 61735001, 61761136002), the Major National Basic Research Program of China (2013CB922401), and the National Key Scientific Instrument and Equipment Development Program (2012YQ140005). The authors are in debt to Yige Lin and Zhen Sun from National Institute of Metrology for providing the CW lasers and the related technical supports.

**References**
1. J. E. Stalnaker, S. A. Diddams, T. M. Fortier, K. Kim, L. Hollberg, J. C. Bergquist, W. M. Itano, M. J. Delany, L. Lorini, W. H. Osakay, T. P. Heavner, S. R. Jefferts, F. Levi, T. E. Parker, and J. Shirley, Appl. Phys. B **89**, 167 (2007).
2. Y. Salvadé, N. Schuhler, S. Lévêque, and S. L. Floch, Appl. Opt. **47**, 2715 (2008).
3. G. Marra, H. S. Margolis, S. N. Lea, and P. Gill, Opt. Lett. **35**, 1025 (2007).
4. J. Millo, R. Boudot, M. Lours, P. Y. Bourgeois, A. N. Luiten, Y. Le Coq, Y. Kersalé, and G. Santarelli, Opt. Lett. **34**, 3707 (2009).
5. C. Li, A. J. Benedick, P. Fendel, A. G. Glenday, F. X. Kärtner, D. F. Phillips, D. Sasselov, A. Szentgyorgyi, and R. L. Walsworth, Nature **452**, 610 (2008).
6. D. Nicolodi, B. Argence, W. Zhang, R. L. Targat, G. Santarelli, and Y. Le Coq, Nat. Photonics **8**, 219 (2014).
7. L. Ma, Z. Bi, A. Bartels, L. Robertsson, M. Zucco, R. S. Windeler, G. Wilpers, C. Oates, L. Hollberg, and S. A. Diddams, Science **303**, 1843 (2004).
8. N. Ohmae, N. Kuse, M. Fermann, and H. Katori, Appl. Phys. Express **10**, 062503 (2017).
9. H. Leopardi, J. D. Rodriguez, F. Quinlan, J. Olson, J. A. Sherman, S. A. Diddams, and T. M. Fortier, Optica **4**, 879 (2017).
10. T. R. Schibli, I. Hartl, D. C. Yost, M. J. Martin, A. Marcinkevicius, M. E. Fermann, and J. Ye, Nat. Photonics **2**, 355 (2008).
11. C. Li, Y. Ma, X. Gao, F. Niu, T. Jiang, A. Wang, and Z. Zhang, Appl. Opt. **54**, 8350 (2015).
12. T. Jiang, A. Wang, G. Wang, W. Zhang, F. Niu, C. Li, and Z. Zhang, Opt. Express **22**, 1835 (2014).
13. S. T. Sørensen, O. Bang, B. Wetzel, and J. M. Dudley, Opt. Commun. **285**, 2451 (2012).
14. Y. Ma, B. Xu, H. Ishii, F. Meng, Y. Nakajima, I. Matsushima, T. R. Schibli, Z. Zhang, and K. Minoshima, Opt. Lett. **43**, 4136 (2018).